\documentclass{ws-ijmpe}
\usepackage[super,compress]{cite}
\begin{document}


\catchline{}{}{}{}{}

\title{Looking for quark saturation in proton and nuclei}

\author{Wei Zhu}

\address{Department of Physics, East China Normal University,
Shanghai 200062, P.R. China\\
wzhu@phy.ecnu.edu.cn}

\author{Rong Wang}

\address{Institute of Modern Physics, Chinese Academy of
Sciences, Lanzhou 730000, P.R. China\\
rwang@impcas.ac.cn}

\author{Jianhong Ruan}

\address{Department of Physics, East China Normal University,
Shanghai 200062, P.R. China\\
jhruan@phy.ecnu.edu.cn}

\maketitle

\begin{history}
\received{Day Month Year}
\revised{Day Month Year}
\end{history}

\begin{abstract}
The quark saturation behavior at low $Q^2$ is shown in a numeric
solution of the DGLAP equation with parton recombination corrections,
which resembles the widely discussed JIMWLK saturation of gluons.
Our calculation suggests that the partonic saturation can be interpreted as
a dynamical balance between the splitting and the fusion processes of partons,
without any other condensation mechanisms added.
The nuclear shadowing saturation at small $x$ resulted from
the proposed quark saturation is also discussed.
\end{abstract}


\ccode{PACS numbers: 12.38.Bx, 12.40.Vv, 13.60.Hb, 25.30.Fj}

\section{Introduction}
\label{intro}

The Jalilian-Marian-Iancu-McLerran-Weigert-Leonidov-Kovner (JIMWLK) equation
\cite{Jalilian-Marian1997prd,Jalilian-Marian1997npb,Weigert2002,Iancu2001npa,Iancu2001plb}
sums the contributions of multi-gluon fusions to
the Balitsky-Fadin-Kuraev-Lipatov (BFKL) evolution
\cite{Lipatov1976sjnp,Fadin1975plb,Kuraev1976,Kuraev1977,Balitsky1978,Balitsky1979}.
An interesting property of JIMWLK equation is a flattish solution of
the unintegrated gluon distribution $\phi_g(x,k_T)$ in the region of
small transverse momentum $k_T<Q_s$,
in which $Q_s$ is commonly referred as the saturation scale.
Assuming that the transition into the saturation region
occurs abruptly near $k_T\sim Q_s$, one can find that
the majority of gluons in this saturation solution
have transverse momentum $k_T\sim Q_s$.
This special solution is speculated as a new state of matter -
the color glass condensation (CGC), which has been discussed massively.
However, the nature of the CGC is still unclear.

Generally, the partonic saturation implies that the occupation of partons
in a fast proton reaches a limit and the number of partons stops growing up.
To estimate the saturation scale simply,
one can apply the following naive picture \cite{Gribov1983}.
The integrated parton distributions $xf(x,Q^2)$ at scale of $Q^2$ is
the number of partons per unit rapidity interval
(the parton rapidity is defined as $y=ln(1/x)$ and $xf(x,Q^2)=dN/dy$).
Therefore the transverse space per parton is $\pi R^2_N/xf(x,Q^2)$
in a nucleon of radius $R_N$, and the cross section for two parton interaction
is estimated to be $C\alpha_s/Q^2$, in which $C$ is a model-dependent factor.
The saturation effect is important when the number of partons per unit of rapidity
times the gluon-gluon interaction cross section approaches the geometric size of the nucleon.
We call it a full occupation limit.
Under this assumption, the corresponding scale $Q_s$ can be estimated by
\begin{equation}
Q_s^2=C\alpha_s\frac{xf(x,Q^2_s)}{\pi R^2_N}.
\label{QSaturation}
\end{equation}
The saturation critical boundary $Q_s(x)$ separates the dilute and
dense partonic systems in the $(x,Q^2)$ plane.
Assuming $xf(x,Q^2)=\pi R^2_NQ^2/(C\alpha_s)$
is required at $Q< Q_s$, the resulting flat distribution of $xf(x,Q^2)$ is
a clear feature of the saturation.
Note that it is difficult to determine the value of $Q_s$ in the experiment
because the transition into saturation may not occur abruptly,
even though $Q_s$ can be given in a detailed calculation.
From above estimation, the following saturation features are given,
which are useful in searching/discriminating the partonic saturation in experiments.
(i) The distribution is strong $Q^2$-dependent at $Q< Q_s$, which is written as
$xf(x,Q^2)\sim \alpha_s^{-1}Q^2$,
while it is weakly $Q^2$-dependent at $Q>>Q_s$
according to the standard linear QCD evolution equations,
which is written as $xf(x,Q^2)\sim \ln (Q/\Lambda_{QCD})$;
(ii) The values of $xf(x,Q^2)$ at $Q<Q_s$ are $x$-independent if
$x$ enters the sea quark dominated region;
(iii) The corresponding unintegrated parton distribution
$k_T\phi(x,k_T^2)$ has a peak at $k_T\sim Q_s$
(the integrated parton distribution $xf(x,Q^2)$ is the $k_T$ integrated
distribution of $\phi(x,k_T^2)$,
and here we have $\phi(x,k^2_T)\simeq\partial xf(x,Q^2)/\partial Q^2\vert_{Q^2=k^2_T}$).

The deep inelastic scattering (DIS) data at small $x$ and low $Q^2$
is an interesting domain for high energy physics, where the partons have a larger
correlation length $\sim 1/Q$ and stronger recombination strength.
Especially, more attention should be paid to small $x$ and low $Q^2$
for searching the parton saturation.
In this paper, we try to look for the possible quark saturation at low $Q^2$.
In our previous work \cite{Chen2014,Chen2014-nuclei}, the parton distributions
in the proton and nuclei are dynamically generated from a extreme low resolution scale
where the nucleon have merely valence quarks
by using a nonlinear QCD evolution equation - the DGLAP equation
with the recombination corrections
\cite{Zhu:npb551,Zhu:npb559,Zhu:hep}.
We find that the produced sea quark distributions present a positive and flat distribution
in the region of small $x$ and low $Q^2$.
It is surprising to find that the saturation behaviors are also shown for the quark distributions,
which will be detailed in Sec. \ref{NonlinearSolution}.

At low $Q^2$ scale, the naive parton distributions partly contribute to
the measured structure functions. The nonperturbative QCD contributions
are not negligible to the structure functions, which make it difficult
to look for the possible quark saturation signature at low $Q^2$.
In order to identify the quark saturation from the experimental data,
we need to subtract the nonperturbative contributions from the measured
structure function $F_2^p(x,Q^2)$ at $Q^2<1GeV^2$. The popular
phenomenological model - the vector meson dominance (VMD) model
\cite{Sakurai1969,Bauer1978,Grammer1978}
is used  to mimic these nonperturbative corrections, which is
described in Sec. \ref{ProtonSaturation}.

In order to confirm the suggested quark saturation for free proton above,
we also investigate the previous experimental data of
nuclear shadowing effect at small $x$ and low $Q^2$.
The nuclear shadowing saturation resulted from the quark saturation in nuclei,
a flat distribution of structure function ratio at small $x$ and low $Q^2$,
is discussed in Sec. \ref{NuclearSaturation}.
Moreover, it is found that the nuclear shadowing effect has different $Q^2$-dependent
behaviors inside and outside of the saturation domain.
More precise experiments at low $Q^2$ are needed to verify the quark saturation.
Discussions and summary are given in Sec. \ref{Summary}.

\section{A solution of quark saturation in nonlinear QCD evolution equation}
\label{NonlinearSolution}

\begin{figure}
\centering
\includegraphics[width=0.55\textwidth]{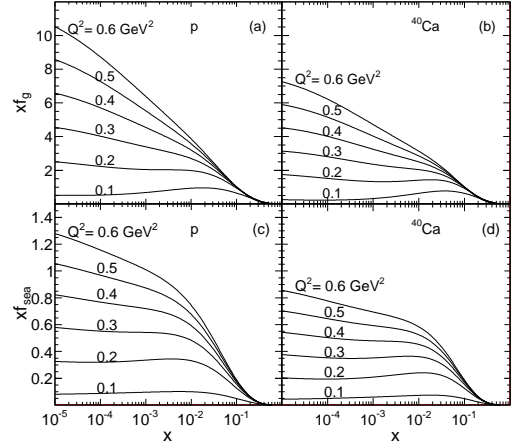}
\caption{Our predicted parton distributions in the low $Q^2$ range,
where the platform-like distributions are found for sea quarks at very low scale.}
\label{fig:1}       
\end{figure}

The physics of parton distributions at low $Q^2$ are rarely discussed.
Nevertheless it is an important and interesting topic as
it is related to the origins of sea quarks and gluons at high $Q^2$.
In our previous work, the gluon and sea quark distributions in the proton
at $Q^2<1 GeV$ are given by the nonlinear evolution equation
\cite{Zhu:npb551,Zhu:npb559,Zhu:hep}, which are shown in Fig. \ref{fig:1}.
Fig. \ref{fig:1}(c) presents one interesting saturation feature of sea quarks.
The sea quark distribution $xf_{sea}(x,Q^2)$ reach a plateau at $x<0.01$
and $0.1 GeV^2<Q^2<0.6 GeV^2$, and the height of the plateau
linearly goes up with $Q^2$ increasing.

\begin{figure}
\centering
\includegraphics[width=0.47\textwidth]{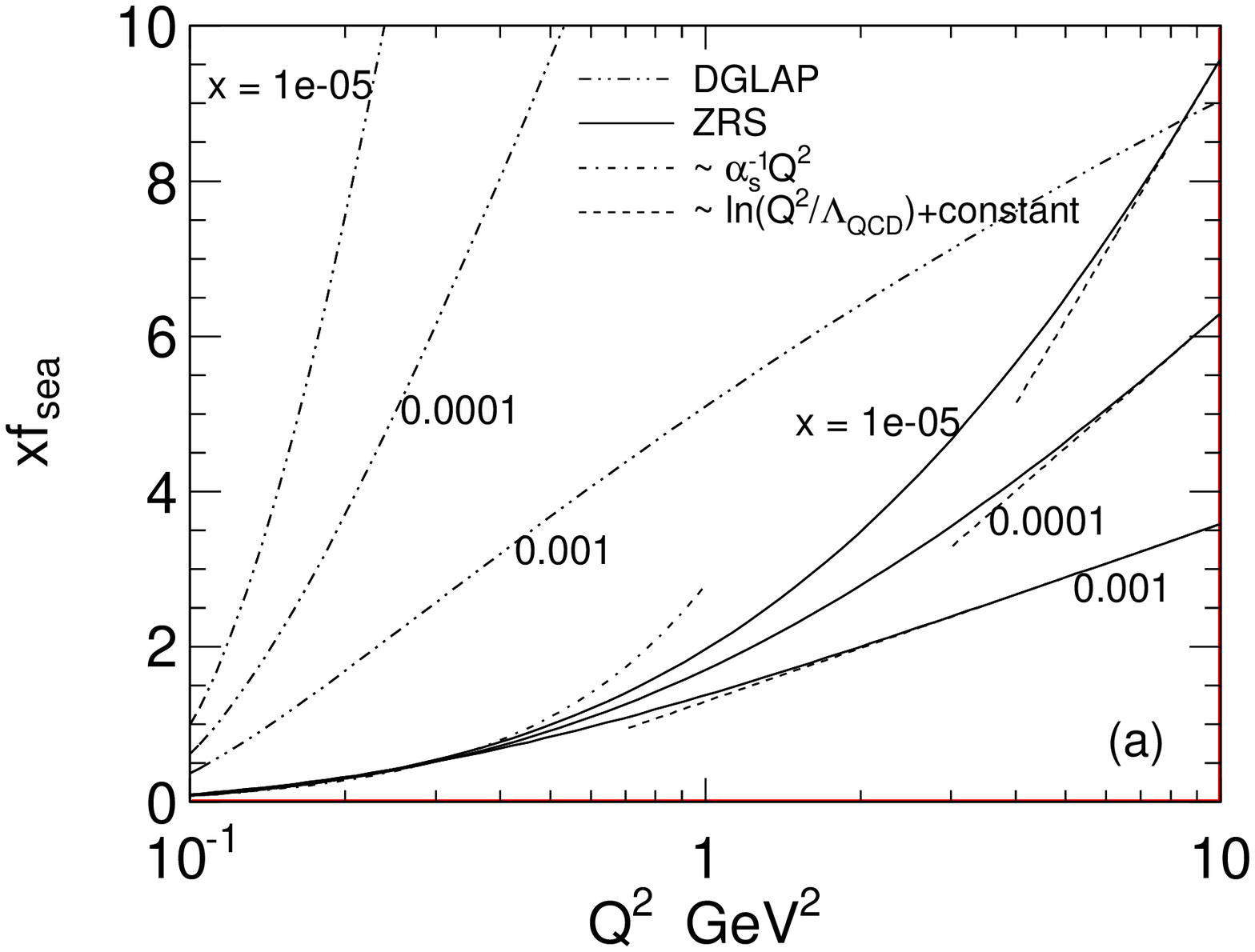}
\includegraphics[width=0.47\textwidth]{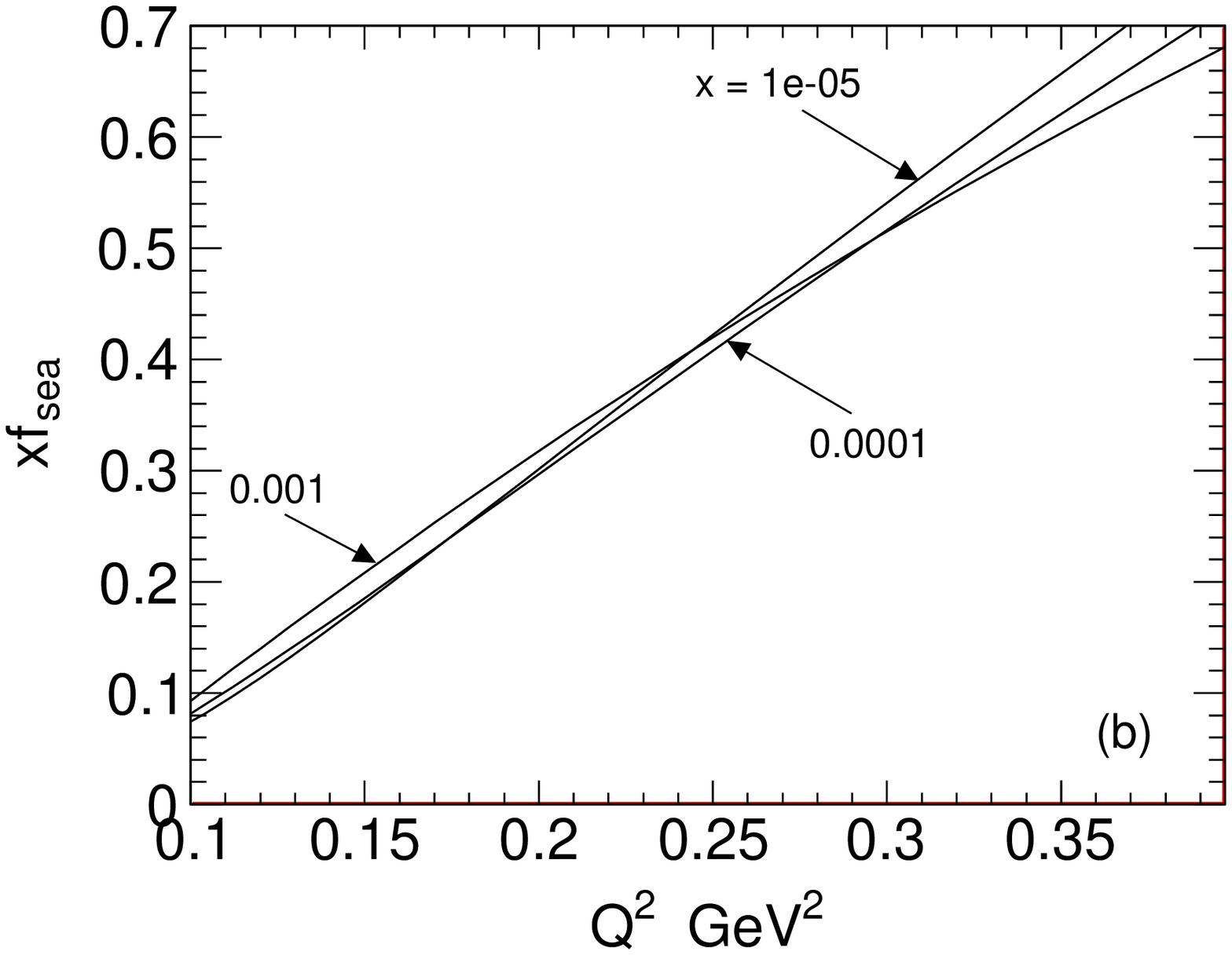}
\caption{Our predicted proton structure functions (solid curves) at
low $Q^2$ and small $x$. The asymptotic curves of the $Q^2$-dependence
of the structure function $F_2$ are shown: function form
$xf_{sea}\sim \alpha_s^{-1}Q^2$ (dash-dot curve) at low $Q^2$ and
function form $xf_{sea}\sim \ln Q^2$ (dash curve) at high $Q^2$.
The dash-dot-dot curves are the solutions of the linear DGLAP
equation. The transition between the two asymptotic behaviors is one
main characteristic of the parton saturation.}
\label{fig:2}       
\end{figure}

Two asymptotic function
$xf_{sea}(x,Q^2)\propto \alpha_s^{-1}Q^2$ and
$xf_{sea}(x,Q^2)\propto \ln Q^2$ are shown together with our
predicted sea quark distribution as a function of $Q^2$ in Fig. \ref{fig:2},
to demonstrate the different $Q^2$-dependence at low and high $Q^2$.
Sea quark distributions $xf_{sea}(x,Q^2)$ generated by the DGLAP
\cite{Altarelli1977,Gribov1972,Dokshitzer1977}
evolution are also shown. The obvious difference between
the result by nonlinear evolution and that by the DGLAP evolution
indicates that the nonlinear effect is substantial in low $Q^2$ range.
It is easy to find the saturation features (i) and (ii) discussed in Sec.
\ref{intro} from our predicted quark distributions. The coincidence of the
predicted parton distributions at different $x$ shown in Fig. \ref{fig:2}(b)
implies a geometrical scaling behavior
$xf_{sea}(x,Q^2)=\psi(Q^2/\Lambda^2_{QCD})$.

\begin{figure}
\centering
\includegraphics[width=0.55\textwidth]{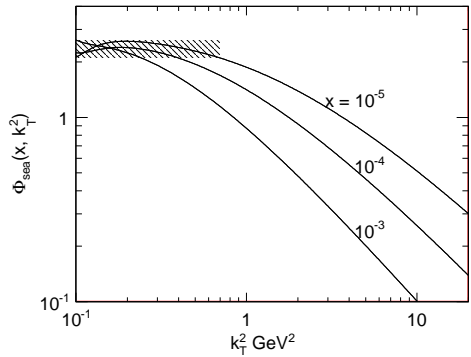}
\caption{Unintegrated sea quark distributions in the proton at low
$Q^2$. The cloud square presents the quark saturation domain.}
\label{fig:3}       
\end{figure}

A more clear demonstration of the quark saturation solution presented
in Fig. \ref{fig:3}, is the unintegrated quark distribution
$\phi_{sea}(x,k^2_T)\simeq [\partial xf_{sea}(x,Q^2)/\partial
Q^2]\vert_{Q^2=k^2_T}$, where the contributions of the Sudakov
factor are neglected at very small $x$ \cite{Kimber2001}.
Note that our resulting curves are determined by the data at high $Q^2$,
which have some errors derived from the experimental errors.
A small uncertainty in $f_{sea}(x,Q^2)$ calculation
results in some obvious undulation in $\phi_{sea}(x,k^2_T)$.
Hence in the geometrical scaling range, the parton distributions at
different $x$ are not ideally equal (Fig. \ref{fig:2}(b)).
In Fig. \ref{fig:3}, the unintegrated quark distributions in the saturation domain
vary in a narrow range (the shadowed block shown in Fig. \ref{fig:3}),
instead of a plateau line predicted by the JIMWLK equation.

\begin{figure}
\centering
\includegraphics[width=0.55\textwidth]{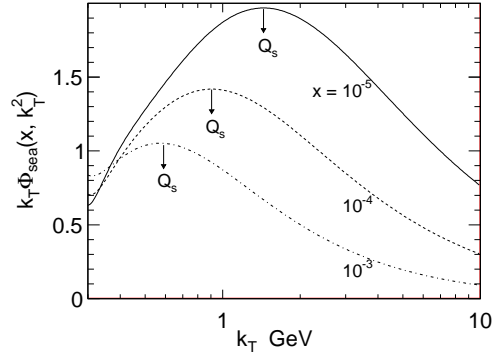}
\caption{Unintegrated sea quark distributions in the proton
(multiplied by the phase space factor of the quark transverse
momentum $k_T$) as a function of $k_T$. The quark momentum accumulation
near $k_T\sim Q_s$ is a characteristic of quark saturation.}
\label{fig:4}       
\end{figure}

\begin{figure}
\centering
\includegraphics[width=0.55\textwidth]{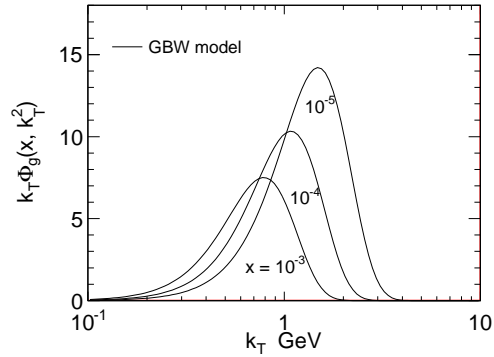}
\caption{A typical unintegrated gluon distribution simulated by the
GBW model \cite{Golec-Biernat1998,Golec-Biernat1999}.
Note that the GBW gluon lacks a $1/k_T$ tail since
the parton evolution is not applied in this model.}
\label{fig:5}       
\end{figure}

Our predicted unintegrated quark distribution $k_T\phi_{sea}(x,k^2_T)$
(multiplied by the phase space factor of the quark transverse
momentum $k_T$) as a function of $k_T$ is shown in Fig. \ref{fig:4},
showing that the major sea quarks have momentum of $k_T\sim Q_s$.
We suggest to call the peak which is of the largest occupation numbers
the quark saturation rather than the condensation. It is also found that
the transition from the normal diluted parton state to the saturation limit
occurs though a rather broad range of $Q^2$, i.e., the transition into
the saturation does not occur abruptly. For reference, similar distributions
for the gluon density in the GBW (Golec-Biernat W\"{u}sthoff) model
\cite{Golec-Biernat1998,Golec-Biernat1999}
are shown in Fig. \ref{fig:5}. This model is
generally used to simulate the JIMWLK gluon saturation.
Note that the gluon distribution by GBW lacks $~1/k_T$ tail.
The reason is that the parton evolution is not treated in this model.

A surprise is that there is no obvious saturation features for the gluon
distribution in the low $Q^2$ range (See Fig. \ref{fig:1}(a)),
although the recombination effect for gluon distribution is much stronger
than that for sea quark distributions as the two leading nonlinear corrections
are both about gluons $P_{gg\rightarrow gg}\sim 1/x$
and $P_{q\overline{q}\rightarrow gg}\sim 1/x$ at small $x$ \cite{Zhu:npb551,Zhu:npb559,Zhu:hep}.
The absence of gluon saturation at low $Q^2$ can be understood as follows.
The main source of gluons near the starting evolution scale is the radiation of valence quarks.
The newly generated gluons from valence quarks
break the possible saturation balance between gluon splitting and gluon combining.
This situation is different from the JIMWLK saturation,
where the gluon radiation from valence quarks
is overwhelmed by the gluon splitting at small $x$.
Away from the starting scale, the number of gluons grow very fast
and dominates at small $x$.
The radiated gluons directly from the valence quarks is relatively of small quantity at high $Q^2$ scale.

The phenomenon of the gluon saturation widely discussed is known as the CGC.
Current theoretical understanding suggests that
the CGC is the collective gluon excitations based on the thought
that the JIMWIK equation sums all orders of gluon fusions.
However, the quark saturation from nonlinear evolution equation
has the similar features of the CGC,
which implies that the partonic (gluon and quark) saturation
is a dynamical balance between strong splitting and fusion processes
rather than a new condensation mater state.
The origin of partonic (gluon and quark) saturation does not need
any condensation dynamics.

\section{Quark saturation shown in the proton structure function at low $Q^2$}
\label{ProtonSaturation}

\begin{figure}
\centering
\includegraphics[width=0.67\textwidth]{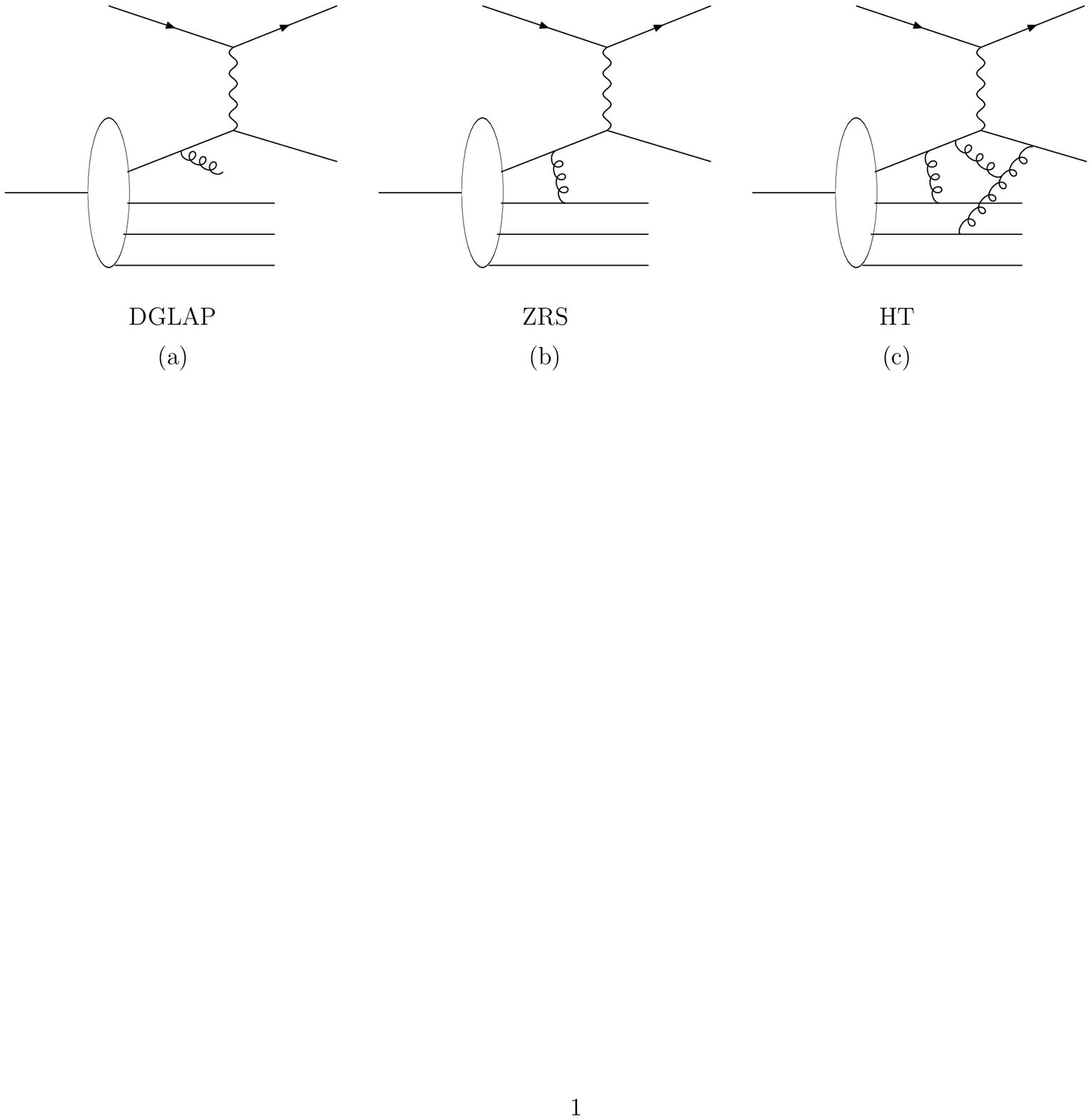}
\caption{DIS processes. (a) The leading twist contributions;
(b) A twist-4 corrections; (c) The high twist contributions, which can
be isolated by using a naive VMD model.}
\label{fig:6}       
\end{figure}

Is the quark saturation only a mathematic solution in a nonlinear QCD evolution equation,
or a real physical existence? To answer this question, we should look for
the observable patterns of the quark saturation in experiments.
Parton distribution functions are extracted from the measured structure
functions, however the nonperturbative QCD components are mixed
with the possible saturation signature at low $Q^2$.
In fact, at low $Q^2$, the multi-parton correlations dominate,
and the inclusive lepton-nucleon cross section is mainly from
the complicate higher twist interactions.
According to the operator product expansion (OPE),
the proton structure function $F_2^p(x,Q^2)$ can be written
as a series in $1/Q^2$,
\begin{equation}
F_2^p(x,Q^2)=F_2^{LT}(x,Q^2)+F_2^{HT}(x,Q^2).
\label{F2LTHT}
\end{equation}
The leading (twist-2) term corresponds to scattering from one single free parton,
while higher twist terms correspond to multi-parton interactions.
Up to date, only a part of higher twist effects, for example, the recombination of initial partons (Fig. \ref{fig:6}(b))
has been calculated perturbatively as the corrections to DGLAP evolution,
which have been discussed in the previous section.
We denote this single parton scattering contribution to the structure function
as $F_2^{DGLAP+ZRS}(x,Q^2)$.
However, we can neither perform nor interpret a partonic calculation of higher twist effects
of the correlations between the initial and finite partons,
since they break the factorization schema.
In a certain kinematic region, some of such higher twist contributions to $F_2^p(x,Q^2)$
appear to be some hadronic interaction phenomena.
For this situation, we may chose a suitable phenomenological model
to describe the corresponding higher twist effects.

\begin{figure}
\centering
\includegraphics[width=0.25\textwidth]{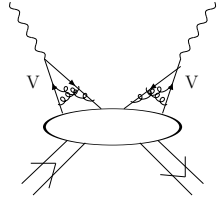}
\caption{The diagram of VMD model.}
\label{fig:7}       
\end{figure}

We try to use the well known Vector Meson Dominance (VMD) model
to mimic the mentioned higher twist corrections above.
The contributions shown in Fig. \ref{fig:6}(c) can not be neglected at low $Q^2$.
The corrections of the quark-antiquark pair is needed,
which interacts with the target like a virtual vector meson
if the transverse momentum $k_\perp \sim Q$ of the quark pair is not large,
and the confinement effects play significant roles.
Since contribution shown in Fig. \ref{fig:6}(c) can not be factorized,
we apply a phenomenological VMD model (Fig. \ref{fig:7}) to calculate it.
Traditionally, the VMD model \cite{Badelek2002epjc,Badelek2002app} was used
to explain the structure function in low $Q^2$ region successfully.
We denote this contribution as $F_2^{VMD}(x,Q^2)$.

In the VMD model,
\begin{equation}
F_2^{VMD}(x,Q^2)=\frac{Q^2}{4\pi\gamma^2_{\rho}}\frac{m^4_{\rho}\sigma_{\rho p}}{(Q^2+m^2_{\rho})^2}.
\label{VMD}
\end{equation}
where $\gamma_\rho$ is the coupling constant between $\rho$ vector meson and proton,
$x$ is a variable defined as $x=Q^2/(s+Q^2-m^2_p)$ rather than the momentum fraction of a parton,
and $s$ is the center-of-mass energy (W) square of the $\gamma p$ collision.
We considered the contributions of $\rho$ meson only, because it is the dominant virtual particle involved.
The cross sections $\sigma_{\rho p}(W)$ is the total cross section of the virtual $\rho$ meson
scattering on the nucleon.

At high energy, the Regge theory \cite{Collins1977} is successful to
parameterize the cross sections $\sigma_{\rho p}$.
The high energy behavior in Regge theory for the total cross section
is expressed as $\sigma_{\rho p}(s)\sim s^{\alpha_P-1}$,
where $\alpha_P=1.0808$ is the intercepts of the Pomeron \cite{Donnachie1998}.
At large $1/x$, there is $\sigma_{\rho p}(x)\sim x^{1-\alpha_P}$.
The elastic scattering dominates in large $x$ region for $F_2^p(x,Q^2)$.
The contributions of Pomeron is fast reduced with $x$ increasing.
We neglect the Regge contribution at large $x$ in this work.
In consequence,
\begin{equation}
F_2^{VMD}(x,Q^2)\simeq B\frac{m^2_{\rho}Q^2}{(Q^2+m^2_{\rho})^2}x^{1-\alpha_P}(1-x)^{20},
\label{VMDAndCutOff}
\end{equation}
where $B$ is a free parameter and determined to be $B=0.4$
from a fit to the experimental data at low $Q^2$.
An arbitrary large power in the factor $(1-x)^{p}(p>>1)$
is applied to suppress the contributions of Pomeron at large $x$.
Note that the contribution of $F_2^{DGLAP+ZRS}(x,Q^2)$ is taken from Ref. \cite{Chen2014},
which is already fixed and determined by the experimental measurements at high $Q^2> 4~ GeV^2$.

More complicated corrections to $F_2^p(x,Q^2)$ at low $Q^2$ are from
the higher order QCD effects ${\cal{O}}(\alpha_{s})$ and the higher
order recombinations. In principle, we need to consider all these
corrections, however it is still worthwhile to look at the leading
corrections at the beginning. Our interest is that if higher order corrections
are small down to some low scale $Q^2(\sim \mu^2)$, then our leading
order analysis of structure function data to $\mu^2$ is acceptable.
Otherwise, our calculation is still important in order to extract the higher
order contributions.

Finally, the proton structure function without higher order QCD effects and
higher order recombinations is written as
\begin{equation}
F_2^p(x,Q^2)\simeq PF_2^{DGLAP+ZRS}(x,Q^2)+F_2^{VMD}(x,Q^2),
\label{DualityFor}
\end{equation}
where $P$ is the probability of inelastic events via bare photon-parton interaction.
As $m_\rho^4/(Q^2+m_\rho^2)^2$ in Eq. (\ref{VMDAndCutOff}) is the probability of the VMD event,
the factor $P$ in Eq. (\ref{DualityFor}) for the single parton scattering is then given by
\begin{equation}
P=1-\frac {m_\rho^4}{(Q^2+m_\rho^2)^2}.
\label{PFormula}
\end{equation}

\begin{figure}
\centering
\includegraphics[width=0.63\textwidth]{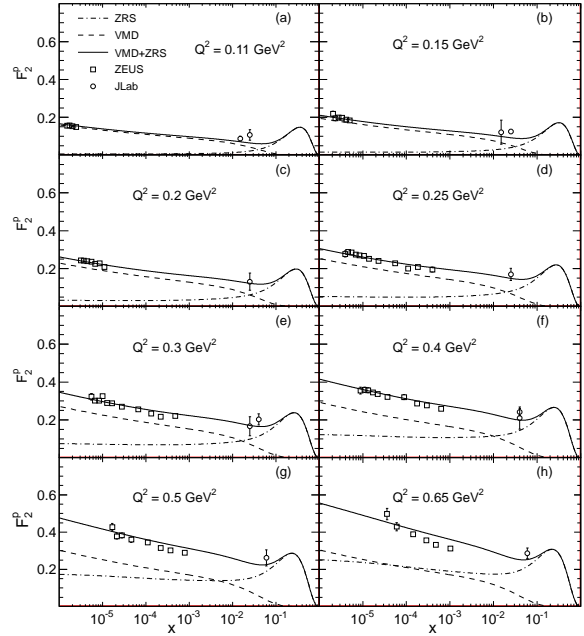}
\caption{The proton structure function $F_2^p(x,Q^2)$ as a
function of $x$ at various $Q^2$ ($<1~GeV^2$).
The contributions of $F_2^{hadron}(x,Q^2)$ and $F_2^{DGLAP+ZRS}(x,Q^2)$
are shown separately.
The data are taken from \cite{ZEUS,JLab,NMC,H1:1996,H1:2001,E665}.
One can find that a flattish quark distribution is important
to the measured structure function.}
\label{fig:8}       
\end{figure}

\begin{figure}
\centering
\includegraphics[width=0.52\textwidth]{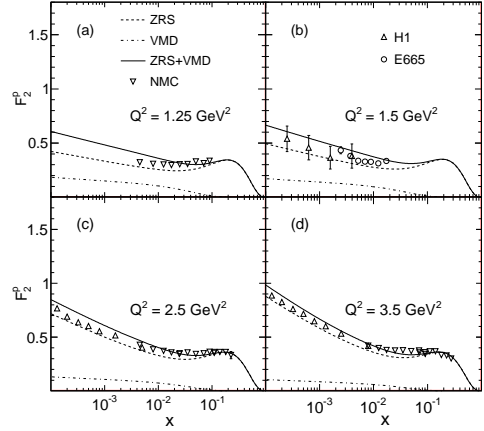}
\caption{Same as Fig. 8 but for $Q^2>1GeV^2$.}
\label{fig:9}       
\end{figure}

The results of our model is compared to experimental measurements
\cite{ZEUS,JLab,NMC,H1:1996,H1:2001,E665}
in Fig. \ref{fig:8} (at $Q^2<1 GeV^2$) and in Fig. \ref{fig:9} (at $Q^2>1 GeV^2$).
The qualities of the fits are good. Our model about the quark saturation
gives consistent results with the observable data.

\section{Quark saturation shown in nuclear shadowing at low $Q^2$}
\label{NuclearSaturation}

The cross section ratios of inelastic muon scattering on nuclei to that on deuterium
at very small $x$ ($0.00002<x<0.25$) and low $Q^2$ down to
$0.1GeV^2$ was measured by Fermilab E665 experiment
\cite{E665:1992,E665:1995}.
The cross section ratio shows a flat distribution at $x< 10^{-3}$.
This nuclear shadowing saturation was confirmed by the later NMC data \cite{NMC:1995,NMC:1991}.
These flat distributions are a sign of the quark saturation.

\begin{figure}
\centering
\includegraphics[width=0.6\textwidth]{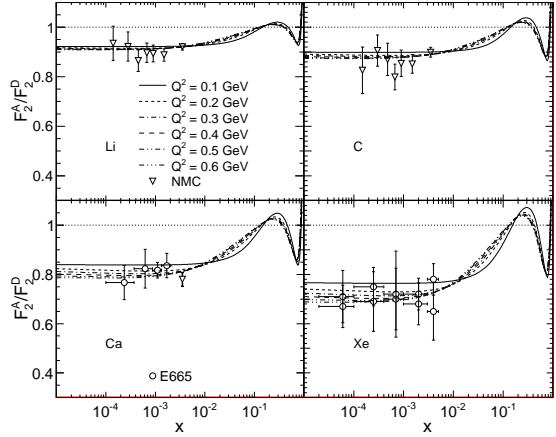}
\caption{Predicted ratios of $F_2^A/F_2^p$ for several nuclei at
$Q^2<1~ GeV^2$ compared to experimental data \cite{E665:1992,E665:1995,NMC:1995,NMC:1991}.
The quark saturation is important for the observed platform-like distribution.}
\label{fig:10}       
\end{figure}

The saturation observed in the nuclear structure function ratio is related to
two different saturation mechanisms. The saturation of
$F_2^{DGLAP+ZRS}(x,Q^2)$ resulted from the quark saturation has been
discussed in Sec. \ref{NonlinearSolution}. The quark saturation feature
is clearly presented for the nuclear quark distributions at small $x$ \cite{Chen2014-nuclei},
which is shown in Fig. \ref{fig:1}(d).
The other mechanism for the saturation of nuclear structure function ratio is due to
the hadron part of the structure function $F_2^{VMD}$.
The underlying process is the multiple-scattering of virtual meson off nuclei.
In the VMD model for the nuclei,
\begin{equation}
F_2^{A,VMD}(x,Q^2)=\frac{Q^2}{4\pi\gamma_{\rho}}\frac{m^4_{\rho}\sigma_{\rho A}}{(Q^2+m^2_{\rho})^2}.
\label{VMDnuclear}
\end{equation}
Using the Glauber eikonal approximation \cite{Glauber1959,Glauber1970},
the relation between the total cross sections $\sigma_{VA}$ and $\sigma_{VN}$ is given by
\begin{equation}
\sigma_{\rho A}=2\int d^2b[1-e^{-\sigma_{\rho p}T_A(b)/2}]\simeq G(R_A/l_V)\sigma_{Vp},
\label{rphoSectionNuclear}
\end{equation}
where $T_A(b)$ is the nuclear thickness function and
\begin{equation}
G(x)=\frac{3}{x^3}\left[(1+x)e^{-x}-1+\frac{1}{2}x^2\right],
\label{GFormula}
\end{equation}
and $l_V$ is the mean free path of the hadronic fluctuation constituent of the virtual photon \cite{Grammer1978}.
It is expected that $l_V$ is much larger than the mean free path of the real vector
meson in the nucleus. Figure \ref{fig:10} shows the results of nuclear structure function ratios
at low $Q^2$ using $l_V=8$ fm.
There is strong kinematic correlation between $x$ and $Q^2$ in the fixed target experiments,
which makes the low $x$ measurements also at low $Q^2$.
Our predicted ratios at $x<10^{-3}$ exhibit a weak dependence on $x$ and $Q^2$,
which are consistent with the experimental data.
However more precise measurements are highly needed.
A wide energy and four-momentum transfer range
of the future Electron-Ion Collider (EIC) would provide
valuable opportunities to search for the quark saturation.

\begin{figure}
\centering
\includegraphics[width=0.6\textwidth]{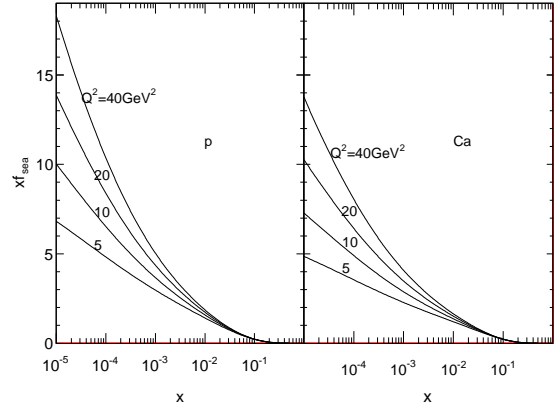}
\caption{Sea quark distributions in proton and calcium at $Q^2=5-40~GeV^2$.}
\label{fig:11}       
\end{figure}

\begin{figure}
\centering
\includegraphics[width=0.55\textwidth]{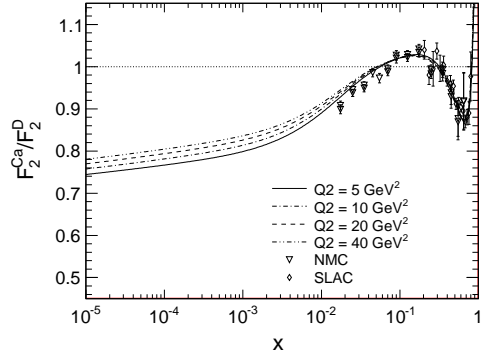}
\caption{Predicted ratios for $F_2^{Ca}/F_2^p$ at $Q^2>>1 ~GeV^2$.
It has different $Q^2$- and $x$-dependence compared to
shadowing saturation at low $Q^2$.}
\label{fig:12}       
\end{figure}

The other possible solution for the structure function ratio saturation
is also considered: the shapes of quark distribution functions in proton and nucleus
are similar towards small $x$ at large $Q^2$ yet not flat.
Our predicted $F_2^p(x,Q^2)$ and $F_2^{Ca}(x,Q^2)$ at $Q^2>>1GeV^2$
are shown in Fig. \ref{fig:11}, where the nonlinear corrections are negligible.
The platform-like distribution is not found at high $Q^2$ (see Fig. \ref{fig:12}).
The $Q^2$- and $x$-dependent behaviors of the nuclear shadowing at high $Q^2$
are different from the nuclear shadowing saturation.

\section{Discussions and Summary}
\label{Summary}

\begin{figure}
\centering
\includegraphics[width=0.45\textwidth]{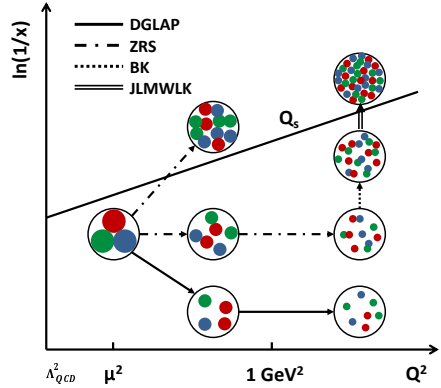}
\caption{A schematic diagram for illustrating the different evolution processes
of parton distributions in different kinematic regions.}
\label{fig:13}       
\end{figure}

The typical parton evolution configuration of the proton in $(x,Q^2)$ space
is shown in Fig. \ref{fig:13}. The starting point of the evolution is the proton
at extreme low $\mu^2$, which is represented as mere three valence quarks.
With the increase of the virtuality of the probe while the value of $x$ is fixed,
we are able to resolve the partons into smaller ones.
This type of evolution is well described by the DGLAP equation
at relatively large $x$, and the DGLAP equation including parton recombination
corrections at small $x$. In the first case, the number of partons
rises logarithmically while their typical size decreases like
$\alpha_s/Q^2$ so that the partons in the proton become more and more dilute.
However in the second case, the number of sea quarks rises as $\sim Q^2/\alpha_s$,
and it keeps in the full occupation situation (splitting and fusion balance),
which leads to the quark saturation.
In a little larger $Q^2$ but very small $x$ range, BFKL evolution
with gluon recombination, namely the BK evolution (with leading
recombination) and JIMWLK evolution (with multi-recombinations)
towards smaller $x$ with $Q^2$ fixed are commonly used.

Different evolution schemes have different kinds of saturations.
The proton in the JIMWLK evolution process leads to the black-disc limit,
for which the gluons are fully overlapped, namely, the JIMWLK saturation.
The three-gluon recombination process of JIMWLK equation plays
an important role in the gluon saturation.
The quark saturation in low $Q^2$ originates from the leading order parton
recombinations with big strong coupling constant.
The saturation for the quark distributions can not be directly probed
in the deep inelastic scattering at low $Q^2$, because the hadronic
component of the probe is mixed with the quark saturation.
The gluon saturation by the JIMWLK equation is the result of
multi-recombination of gluons at high $Q^2$,
which is also can not be observed directly by the electro-magnetic probe.
It is very interesting that these two kinds of saturation have similar features,
which suggests that both the quark saturation and the gluon saturation
comes from the dynamic balance between the parton
recombination and parton splitting.

In summary, a numeric solution of the DGLAP equation with parton recombination corrections
at low $Q^2$ shows the quark saturation, which likes the JIMWLK
saturation of gluons. This calculation shows that the partonic
saturation is the result of the dynamical balance between parton
splitting and fusion in the context of perturbative QCD theory.
The observed nuclear shadowing saturation interpreted as the quark saturation
is discussed. We argue that the nuclear shadowing saturation measured
by E665 and NMC collaborations is the sign of the quark saturation.

\end{document}